# Properties of Nonequilibrium Steady States: a Path Integral Approach

**E.G.D. Cohen**

Laboratory of Statistical Physics, The Rockefeller University, 1230 York Avenue, New York, NY 10065, USA

Abstract

A number of properties of systems in a nonequilibrium steady state (NESS) are investigated by a generalization of the Onsager-Machlup (OM) path integral approach for systems in an equilibrium state (ES). A thermodynamics formally identical to that in an ES can be formulated, but with definitions of work and heat as those needed to maintain the NESS. In this approach, the heat plays a crucial role and is directly related to the different behavior of a system's forward and backward paths in time in an appropriate function space. However, an ambiguity in the choice of the time-backward path corresponding to a given time-forward path prevents a unique general formal theory for systems in a NESS. Unique unambiguous physically acceptable physical results for a system in a NESS appear to be obtainable only after specifying the physical nonequilibrium parameters, which define a system in a NESS as part of a larger system. NESS systems are therefore fundamentally different from those in an ES. Furthermore, an example is given for a particular system that the fluctuations of a system in a NESS behave in many respects very differently from those in a system in an ES.

**Contents:**

1. Introduction

This paper is meant to be an overview of three previous papers [1-3], which apply the OM approach to study fluctuations of systems in an ES [4-5], to fluctuations in systems in a NESS. The paper is not intended to be a survey paper of the literature of all the work done on systems in a NESS; a critical comparison is intended for a future paper [6]. For details and more extensive references, we refer to the original papers.

We will start with a brief discussion of some relevant features of the ES. This in order to be able later to point out what are, in our opinion, the fundamental differences in principle between this state and the NESS, at least from the point of view adapted in the generalized OM theory of the NESS, presented here.

By far the best known and understood steady state is the ES. For a closed system in an ES in contact with a heat bath at temperature $T$, no work has to be done on the system nor does any heat have to be applied to or removed from the system in order to maintain the system in its ES.

On the other hand, microscopically, in terms of the particles out of which the system consists, their (thermal) motion causes fluctuations around the values of the macroscopic (i.e.: average) properties of the system. Thus although there is on average no heat exchange with the environment there will be all the time heat flows into and out of the system so that heat fluctuations have to be considered. However, since no work is done on the system, no work fluctuations have to be considered. This is the case for the OM theory of systems in an ES.

For the macroscopic properties of systems in an ES, a thermodynamic description has been developed culminating in the first (energy conservation) and the second (entropy) laws, connecting heat, work, and internal energy. These laws and their many consequences have been able to describe all macroscopic properties observed in an ES, including phase transitions and quantum effects.

The ES can be considered as a special case of the class of steady states, which also includes nonequilibrium steady states (NESS). Up until very recently there has been no thermodynamic description of systems in a NESS like that for an ES. As will be explained later this appears to be due in essence because a system in a NESS is not a simple system like one in an ES coupled to a heat bath to maintain its ES, but it is connected to other systems needed to maintain its NESS. In fact, if a system is in a NESS, there are nonequilibrium parameters characterizing the nature and the strengths of the couplings of that system (of interest) to other systems, which together define the NESS of the system.

One of the results of this article is an attempt to formulate a thermodynamic description as well as a fluctuation theory for systems in a NESS, using the path integral approach of Onsager and Machlup (OM) for a system in an ES [4,5]. In doing so, some new features of a NESS emerge, which are fundamentally different from those in an ES. As a consequence, the NESS appears to be in some sense a new kind of state, different from an ES. The study of the fluctuations in a NESS also allows us in particular to make a connection with the fluctuation relations for work and heat derived in the last fifteen years for systems with deterministic [7,8] as well as stochastic [9,10] dynamics.

Before we use the OM path integral approach to a theory of fluctuations for a system in an ES for a system in a NESS, we very briefly sketch the OM theory for fluctuations in a system in an ES.

For such systems, OM considered the relaxation *back* to equilibrium of fluctuations $a_j(j=1,.....,n)$ – with zero averages $\bar{a}_j = 0$ – of local physical quantities $A_j$, *away* from equilibrium. This relaxation process was assumed to be described by a linear Langevin equation of the form:

$$\sum_{k=1}^{n} \left[ m_{jk}\ddot{a}_k(t) + r_{jk}\dot{a}(t) + s_{jk}a_k(t) \right] = \zeta_j(t) \tag{1}$$

Here $\zeta_j(t)$ is a Gaussian random noise, $r_{jk}$ is a matrix representing the (dissipative) linear laws of Irreversible Thermodynamics or equivalently the constitutive equations between fluxes (currents) and forces (gradients) in hydrodynamics, while $m_{jk}$ and $s_{jk}$ are matrices representing the first non vanishing (second order) terms in an expression of the local entropy of the system in powers of the fluctuations $\{a_j\}$ and their time derivatives $\{\dot{a}_j\}$, respectively.

We remark that our generalization of the OM theory uses the observation that OM do not use in the development of their theory the specific nature of the $a_j$, as fluctuations of (extensive) local quantities. Therefore one can interpret $a_j$ also as the position $x_j$ of a Brownian particle in a fluid subject to friction, characterized by a matrix $r_{jk}$ and harmonic forces, characterized by a matrix $s_{jk}$.

The generalization to a NESS proceeds as follows [1]. We first note that in a NESS, the averages $\bar{a}_j$ will no longer vanish, i. e., $\bar{a}_j \neq 0$. Since the $\bar{a}_j = 0$ in the ES, the $\{\bar{a}_j\}$ characterize the magnitude of the deviations from equilibrium. Replacing then in (1) the $a_k$ by $(a_k - \bar{a}_k)$, one obtains for $(j=1,....,n)$ the linear Langevin equation:

$$\sum_{k=1}^{n} \left[ m_{jk}\ddot{a}_k(t) + r_{jk}\dot{a}(t) + s_{jk}a_k(t) - s_{jk}\bar{a}_k(t) \right] = \zeta_j(t), \tag{2}$$

where the constant $\sum_k s_{jk}\bar{a}_k(t)$ incorporates the NESS of the system. Although mathematically the difference between (1) and (2) is only a constant, i.e. trivial mathematically, this difference will turn out to be major physically, since it leads to a number of physical properties of the OM theory for systems in a NESS fundamentally different from those in an ES. We will mention three of the most interesting of those properties here.

First, a thermodynamics formally similar to that for systems in an ES can be formulated for systems in a NESS, by using appropriate definitions of work, heat and internal energy. The NESS heat and work are defined as those needed to maintain the NESS: they vanish in an ES.

Second, work distribution functions and a number of fluctuation relations can be derived for systems in a NESS and their dependence on the initial state of such systems, not restricted to either an ES or a NESS initial condition, as was hitherto the case.

Third, ambiguities present themselves, however, in the choice of the work, heat and internal energy for a system in an NESS and which enter also into the thermodynamics of a NESS. The number of ambiguities is directly related to the number of nonequilibrium parameters, mentioned above, i.e. to the complexity of the system.

These multiple possible choices all lead formally to the same thermodynamic laws, but it is not possible to make, in general, a unique choice for them, valid for all systems in a NESS, as is, of course, the case for systems in an ES. In fact, only *after* one has defined the system concretely, a unique choice can be made. For eight concrete models considered in [3], the physically correct definitions of heat, work and energy required one choice for one half of them, while the other half required a different choice to obtain physically acceptable results. The origin of this ambiguity is rooted in an ambiguity in the possible choice of the backward (time reversed) path corresponding to given forward (in time) path in the path integral approach to systems in a NESS. This is due to the presence of physical *nonequilibrium parameters*, which define the NESS system considered and whose parity with respect to time reversal is relevant to determine the appropriate backward path corresponding to a given forward paths as is the case in the models considered in [3].

We now give a brief outline of the content of this paper. In section 2, in order to elucidate various features that distinguish the NESS from the ES, basic Langevin equations for a simple exactly soluble model of a dragged Brownian particle in a harmonic potential are given, as well as a definition of the work to maintain the system in a NESS. This model is used throughout the paper to illustrate various results. In section 3, the path integral approach to obtain the properties of fluctuations via transition probabilities and a (stochastic) Lagrangian is discussed. In section 4, the computation of the work distribution for the dragged Brownian particle model is outlined, from which in section 5 an asymptotic work fluctuation theorem is given, valid for all initial conditions of the NESS system. In section 6 properties of the work fluctuations for the model of section 2 for finite times, which then depend on the initial state, are sketched. The existence of a critical mass is shown to exist, beyond which the fluctuations change from purely decaying to propagating as well. In section 7 the ambiguity in the choice of the time reversed backward path to a given forward path in a system in a NESS is discussed in a general setting. The first and the second laws of a NESS thermodynamics are formulated, which reflect this ambiguity. The necessity to specify the NESS system completely, before a unique choice of the proper physical quantities of such a system can be made, is discussed. In section 8, transient fluctuation relations are derived from two nonequilibrium detailed balance relations for two different kinds of work for the dragged Brownian particle model: the mechanical work $W$, and the friction work $R$, introduced in equations (15) and (41), respectively. While the former obeys both a transient and an asymptotic fluctuation relation, the latter obeys only a transient fluctuation relation. In section 9 some comments and open questions are discussed.

2. <u>Illustration on an exactly soluble model</u>

Our theory can best be illustrated by using a simple system: that of a Brownian particle, confined in its motion by a laser-induced harmonic potential, which is dragged by an external force with constant velocity $\upsilon$ through a fluid (heat reservoir) in an ES at temperature $T$ [1]. Starting the Brownian particle at the initial time $t=0$ at the bottom of the harmonic potential, it will, after a transient motion in the fluid, be dragged to a stationary position in the harmonic potential, where the friction force of the fluid on the particle resisting the dragging motion on the one hand and the harmonic force pulling the particle back to its initial (lowest energy) position on the other hand, balance (cf. fig. 1). More precisely, the system of interest, which consists here of the Brownian particle in a harmonic potential [11-13], has then reached a NESS, in which the Brownian particle is on average at a stationary position in the potential, which moves with constant velocity $\upsilon$ forward, say, in the fluid. The fluctuations considered in this system are those of the position of the Brownian particle around this stationary position. The nonequilibrium parameter for the system is the dragging velocity $\upsilon$, which vanishes for an ES, where the Brownian particle is at the bottom of the harmonic potential.

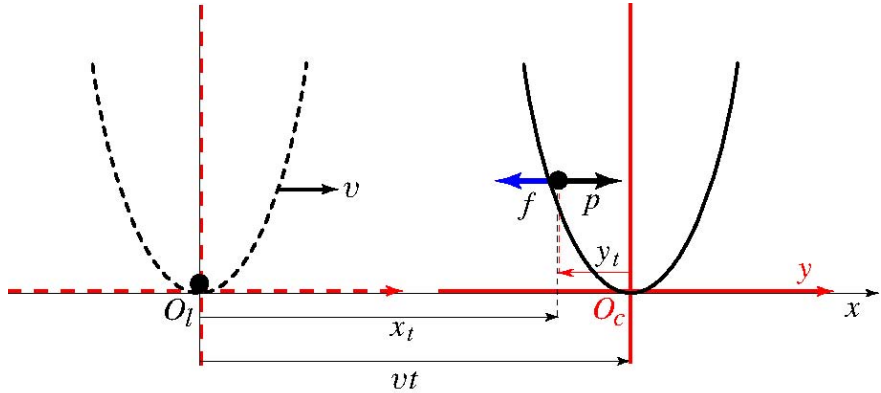

**Figure 1** Schematic illustration of a Brownian particle (black dot), confined in its motion by a laser induced harmonic potential (black parabolas), which is dragged through a fluid with a constant velocity $\upsilon$. Here $x$ with origin $O_l$ (black) and $y$ with origin $O_c$ (red) refer to the axes and origins of the laboratory (*l*) and comoving (*c*) frames, respectively, which are in the direction of $\upsilon$. The particle position is at $x_t(y_t)$ in the laboratory (comoving) frame at time $t$, respectively, related by $y_t = x_t - \upsilon\, t$. The figure shows the position of the particle $y_t$, after a characteristic relaxation time $\tau_r = \alpha/\kappa$ (cf (3)), at the position $y_t = y_{NESS} = -\upsilon\tau_r$, when the system has reached a NESS and where the friction force $f$ (blue) and the mechanical force $p$ (black), due to the harmonic potential on the particle, balance.

Mathematically the motion of the particle can be described by a Langevin equation of the form:

$$m\ddot{x}_t = -\alpha\dot{x}_t - \kappa(x_t - \upsilon\, t) + \zeta_t \qquad (3a)$$

or

$$\tau_m \ddot{x}_t = -\dot{x}_t - \frac{1}{\tau_r}(x_t - \upsilon\, t) + \frac{1}{\alpha}\zeta_t. \qquad (3b)$$

Here $\tau_m = \frac{m}{\alpha}$ and $\tau_r = \frac{\alpha}{\kappa}$ are, respectively, a relaxation time due to the finite mass $m$ of the Brownian particle and a characteristic relaxation time for the decay of the Brownian particle to equilibrium when the mass of the Brownian particle is neglected, i.e. for the so-called overdamped Langevin equation

The first term on the right hand side (r.h.s.) of (3) is the friction force, where α is the friction coefficient between the particle and the fluid, the second term on the r.h.s. of (3) is the harmonic force exerted on the particle at time $t$, which is determined by the elongation of the harmonic "spring" $x_t - \upsilon t$ and has strength $\kappa$ (the harmonic spring constant). The third term on the r.h.s. of (1) is due to the force exerted by the thermal motion of the fluid particles on the Brownian particle, causing its Brownian motion, characterized by a Gaussian random white noise function $\varsigma_t$ with:

$$\langle \varsigma_t \rangle = 0 \text{ and } \langle \varsigma_{t_1} \varsigma_{t_2} \rangle = \left(\frac{2\alpha}{\beta}\right) \delta(t_1 - t_2), \tag{4}$$

where $\beta = \frac{1}{k_B T}$, with $k_B$ Boltzmann's constant.

The friction and the noise terms together represent the effect of the fluid on the Brownian particle. Changing one will therefore require a change in the other. These two terms also give the heat produced in the system by the motion of the Brownian particle through the fluid.

We note that the nonequilibrium parameter, characterizing, the NESS of the system, is for this model the dragging velocity υ. As for all systems in a NESS, the corresponding OM results for the same system in an ES are obtained for vanishing nonequilibrium parameters i.e. for υ = 0 here.

The presence of a constant dragging velocity $\upsilon$ leads to the existence here of two inertial frames of reference : a laboratory and a comoving frame. The two frames are illustrated in figure 1. The position of the Brownian particle $x_t$ in the laboratory frame at time $t$ is related to that in the comoving frame $y_t$ by:

$$y_t = x_t - \upsilon t, \tag{5a}$$

so that

$$\dot{y}_t = \dot{x}_t - \upsilon \text{ and } \ddot{y}_t = \ddot{x}_t. \tag{5b}$$

It is often convenient to use, instead of the Langevin equation (1) in the laboratory frame, the corresponding equation in the comoving frame which has a (constant) velocity $\upsilon$ with respect to the laboratory frame:

$$m\ddot{y}_t = -\alpha \dot{y}_t - \alpha \upsilon - \kappa y_t + \zeta_t, \tag{6a}$$

or

$$\tau_m \ddot{y}_t = -\dot{y}_t - \upsilon - \frac{1}{\tau_r} y_t - \frac{1}{\alpha} \zeta_t. \tag{6b}$$

We remark that the equations (5a) and (5b) allow to transform any expression in terms of the coordinates $x_t$ in the laboratory frame into one in terms of the coordinates $y_t$ in the comoving frame. We also note that the equations (3) and (6) are not identical, which implies a lack of Galilean invariance of the Langevin equation due to the presence of the fluid. Furthermore the fluid is accounted for in the Langevin equation (3) by two terms: one, causing the friction $-\alpha \dot{x}_t$ of the Brownian particle in the fluid when it is forced to move backwards, if the harmonic potential is dragged forward and another, by the incessant "random" collisions of the fluid molecules with the Brownian particle as expressed by $\zeta_t$.

Before we discuss the work fluctuations in this model – for other models we refer to [1-3] - we will give, for later use, a more general formulation of the equation (3), by replacing the harmonic force in (3) by a general mechanical force $F$. Then equation (3a) becomes:

$$m\ddot{x}_t = -\alpha \dot{x}_t - F(x_t, t; \mu) + \zeta_t, \tag{7}$$

where $\mu$ represents in general a nonequilibrium parameter, characterizing the NESS of the system. We have restricted ourselves in this paper as well as in [1-3] to forces linear in $x_s$, so that:

$$\langle F(x_t, t; \mu) \rangle = F(\langle x_t \rangle, t; \mu), \tag{8}$$

where the brackets $\langle \cdots \rangle$ indicate an average over the Gaussian probability distribution (4) of the $\zeta_t$. We remark that this linearity assumption of $F$ is only used – in this paper and in [1-3] – for a derivation of the second law of thermodynamics, although this law should, of course, be valid for any $F(x_t,t;\mu)$; otherwise it is not used in this paper.

3. Fluctuations from Transition Probabilities

As said before, in order to discuss the fluctuation properties of the NESS we generalize the OM path integral approach for an ES to a NESS. We start by defining for the NESS the transition probability $F(x_t\ t\,|\,x_{t_0}\ t_0)$ that, the Brownian particle jumps from an initial position $x_{t_0} \equiv x_0$ at time $t_0$ to a final position $x_t$ at time $t$. Then an initial distribution of the particle $f(x_0,t_0)$ will change to a final distribution $f(x_t,t)$ according to:

$$f(x_t,t) = \int dx_0\ F(x_t\ t\,|\,x_0\ t_0) f(x_0,t_0). \tag{9}$$

Using the functional integral approach of OM[1], one can write the transition probability $F(x_t\ t\,|\,x_{t_0}\ t_0)$ of the Brownian particle from the initial state $x_0$ to a final state $x_t$ as:

$$F(x_t\ t\,|\,x_0\ t_0) = \int_{x_0}^{x_t} \mathcal{D} x_s\ \exp\left[\int_{t_0}^{t} ds\, L(X_s;\upsilon)\right]. \tag{10}$$

Here we use in general $X_s \equiv (\ddot{x}_s, \dot{x}_s, x_s, s)$ as a short hand for the complete dependence of a Lagrangian $L(X_s;\upsilon)$ on $x_s$ and its first two time derivatives and $s$. $L(X_s;\upsilon)$ can be considered as a "stochastic" Lagrangian describing the stochastic dynamics of the Brownian particle based on the Langevin equation, in a somewhat similar way as a dynamical Lagrangian describes the deterministic dynamics of a particle in Mechanics. For the dragged particle model in particular we have that $L(X_s; \upsilon)$ is given by:

$$L(X_s; \upsilon) = -\frac{1}{4D} \left[ \tau_m \ddot{x}_s + \dot{x}_s + \frac{1}{\tau_r} x_s - \frac{s}{\tau_r} \upsilon \right]^2. \tag{11}$$

---

[1] This paper is based on the notion of a functional $\mathcal{F}(\{x_s\})$, short for $\mathcal{F}(\{x_s\}_{s\in[t_0,t]})$, where the function $\{x_s\}_{s\in[t_0,t]}$ describes a particular path from $x_0, t_0$ to $x_t, t$ in function space. $\mathcal{F}(\{x_s\})$ is therefore a function of a function or a functional. $\int_{x_0}^{x_t} \mathcal{D} x_s$ is a functional integral, indicating an integration over *all* paths $\{x_s\}_{s\in[t_0,t]}$ from $x_0, t_0$ to $x_t, t$, where the integral can be defined as a limit of discrete time steps along the path $\{x_s\}_{s\in[t_0,t]}$ (cf [1,4]).

In (11), $D$ is the Einstein diffusion coefficient associated with the Brownian particle's (stochastic) motion through the fluid, given by:

$$D = \frac{1}{\alpha \beta}. \tag{12}$$

We note that in the representation (9) of $F(x, t \mid x_0, t_0)$ the functional $\exp\left[\int_{t_0}^{t} ds\, L(X_s; \upsilon)\right]$ is proportional to the probability functional $P(\{x_s\})$ associated with the occurrence of the path $\{x_s\}_{s \in [t_0, t]}$ in function space. In fact, using the Langevin equation (3) as well as the Gaussian distribution of the noise (4), one obtains immediately a distribution for the probability functional $P(\{x_s\})$:

$$P(\{x_s\}; \upsilon) = C_x \exp \int_{t_0}^{t} ds\, L(X_s; \upsilon), \tag{13}$$

where $C_x$ is a normalization constant and $L(X_s; \upsilon)$ is given by (11).

The path integral average of any functional $\mathcal{F}(\{x_s\})$ over all possible paths $\{x_s\}_{s \in [t_0, t]}$, with the initial condition at $t_0$ of $x_{t_0} \equiv x_0$ and $\dot{x}_{t_0} = \dot{x}_0$ to the final condition at $t$ of $x_t$ and $\dot{x}_t$ is defined by:

$$\langle\langle \mathcal{F}(\{x_s\}) \rangle\rangle_t \equiv \iint dx_0 d\dot{x}_0 \iint dx_t d\dot{x}_t \int_{(x_0, \dot{x}_0)}^{(x_t, \dot{x}_t)} \mathcal{D} x_s\, \mathcal{F}(\{x_s\})\, P(\{x_s\})\, f(x_0, t_0). \tag{14}$$

Here, as well as the rest of the paper, the dependence of $\langle\langle \mathcal{F}(\{x_s\}) \rangle\rangle_t$ and $P$ on $\upsilon$ has been suppressed as well as that of $\langle\langle \cdots \rangle\rangle_t$ on $t_0$.

The necessity to integrate over both the initial state as well as over the final state is due to the stochastic nature of the dynamics for the paths $\{x_s\}_{s \in [t_0, t]}$ because, unlike for deterministic dynamical systems, for stochastic dynamics, the final state is not uniquely determined by the initial state.

### 4. Work distribution for dragged Brownian particle

Due to the lack of Galilei invariance of the Langevin equation, the mechanical work $W_l$ to maintain the NESS in the laboratory frame ($l$) differs from that in the comoving frame ($c$), $W_c$. For:

$$W_l = \int_{t_0}^{t} ds \left[ -\kappa\left(x_s - \upsilon s\right) \right] \upsilon, \qquad (15a)$$

while

$$W_c = \int_{t_0}^{t} ds \left[ -\kappa y_s - m\ddot{y}_s \right] \upsilon, \qquad (15b)$$

so that with (4),

$$W_c - W_l = \int_{t_0}^{t} ds \left( -m\ddot{y}_s \right) \upsilon. \qquad (15c)$$

Here $-m\ddot{y}_s$ is a d'Alembert-like force acting in the comoving frame to keep the Brownian particle on average at a fixed position in this frame [2] (fig. 1). This work difference in the two frames is physically due to the kinetic energy difference

$\Delta K_l = \frac{1}{2}m\dot{x}_t^2 - \frac{1}{2}m\dot{x}_0^2$ in the laboratory frame and $\Delta K_c = \frac{1}{2}m\dot{y}_t^2 - \frac{1}{2}m\dot{y}_0^2$ in the comoving frame, respectively, leading with (5b) to (cf [1,2]):

$$\Delta K_c - \Delta K_l = \int_{t_0}^{t} ds \left( -m\ddot{y}_s \right) \upsilon = \int_{t_0}^{t} ds \left( -m\ddot{x}_s \right) \upsilon. \qquad (15d)$$

It is convenient to introduce a parameter $\vartheta$ labeling the frames, so that $\vartheta = 1 (0)$ refer to the laboratory (comoving) frame, respectively. Then for both frames one has in general:

$$W(\{x_s\}) = \int_{t_0}^{t} ds\, W(X_s) = \int_{t_0}^{t} ds[-\kappa(x_s - \upsilon s) - (1-\vartheta)\, m\, \ddot{x}_s]\, \upsilon. \qquad (16)$$

where in (16), as in the rest of the paper, the parameter $\vartheta$, has been suppressed in $W(\{x_s\})$. The distribution function for the dimensionless work $\beta W(\{x_s\})$ i.e.: the probability that $\beta W(\{x_s\})$ has the value $W$, is given by:

$$P(W,t) = \langle\langle \delta(W - \beta W(\{x_s\})) \rangle\rangle_t. \tag{17}$$

$P(W,t)$ can be obtained from (17) by taking a Fourier transform of $P(W,t)$ and evaluating its path integral average $\langle\langle \cdots \rangle\rangle_t$ defined in (14). This involves finding first the optimal[2] $\{\tilde{x}_s\}_{s \in [t_0, t]}$ path that maximizes a modified Lagrangian [2]:

$$L_w(X_s; \upsilon) = L(X_s; \upsilon) + \lambda \beta W(\{x_s\}), \tag{18}$$

where $L(X_s; \upsilon)$ is given by (11), $W(\{x_s\})$ by (16) and $\lambda$ a Lagrangian multiplier expressing the constraint (17)[2][3].

The optimal path $X_s^*$ for (18) follows from the condition:

$$\delta \int_{t_0}^{t} ds\, L_w(X_s^*; \upsilon) = 0, \tag{19}$$

with the boundary condition that $x_{t_0}^* = x_0, \dot{x}_{t_0}^* = \dot{x}_0$ and $x_t^* = x_t, \dot{x}_t^* = \dot{x}_t$. This leads, as in mechanics, to an Euler-Lagrange equation for $L_w$ of (18), which reduces for our model to a simple linear fourth-order differential equation:

$$\tau_m^{\,2} \ddddot{\tilde{x}}_s^* - (1 - 2\frac{\tau_m}{\tau_r}) \ddot{\tilde{x}}_s^* + \frac{1}{\tau_r} \tilde{x}_s^* = 0, \tag{20a}$$

where $\tilde{x}_s^*$ is defined by:

$$\tilde{x}_s^* = x_s^* - \upsilon s + (1 - 2\lambda)\,\upsilon \tau_r. \tag{20b}$$

---

[2] The terminology used in this paper conforms with that used in the mathematical literature (see e.g. [14]) but differs from that used in [1-3]. Therefore, what was called there the *most probable* or *most contributing* path – which follows from the condition that the time integral over the Lagrangian with fixed initial and final time is a minimum – is here called the *optimal* path. What was called in [1-3] the *average* position – which follows from an average linear Langevin equation, or also from the condition that the Lagrangian itself is a maximum – is here also the *most probable* position. This then holds not only for systems described by a linear Langevin equation, but also by a non-linear Langevin equation.

[3] We note that for $\lambda = 0$, $L_w(X_s; \upsilon)$ reduces to $L(X_s; \upsilon)$, the Lagrangian for the NESS of the system; for the nonequilibrium parameter $\upsilon = 0$, the NESS reduces to an ES, in which, as said before, no work is done.

Considering solutions of (20) of the form $\tilde{x}_s^* \sim \exp(\nu s)$, one obtains a factorizable equation for $\nu$ leading to four possible values of $\nu$: $\nu = \nu_+, \nu_-, -\nu_+, -\nu_-$ where:

$$\nu_\pm = \left[1 \pm \sqrt{1 - 4\frac{\tau_m}{\tau_r}}\right]. \tag{21}$$

The general solution of (20) is then a superposition of these four special solutions in terms of $\pm \nu_\pm$

Next, the real path $\{x_s\}_{s \in [t_0, t]}$ is a sum of the optimal path $\{x_s^*\}_{s \in [t_0, t]}$ and its complement $\{\Delta x_s = x_s - x_s^*\}_{s \in [t_0, t]}$ and obeys the same boundary conditions as (19).

Using this, one has[4]:

$$\int_{t_0}^t ds\, L_w(X_s; \upsilon) = \int_{t_0}^t ds\, L_w(X_s^*; \upsilon) - \frac{1}{4D}\int_{t_0}^t ds [\Delta \dot{x}_s + \frac{1}{\tau_r}\Delta x_s + \tau_m \Delta \ddot{x}_s]^2. \tag{22}$$

Here the last term on the r. h. s. of (22), involving the deviation $\Delta x_s$ of the optimal path $\{x_s^*\}$ from the real path $\{x_s\}$, takes care of the proper normalization of the distribution function and can henceforth be omitted [2]. The first term on the r. h. s. of (22) gives, after some algebra, an explicit Gaussian expression for $P(W,t)$ in the long time limit for *any* initial condition $f(x_i, \dot{x}_i, t_0)$ [2].

### 5. Asymptotic Work Fluctuation Theorem

For $t \to \infty$ the asymptotical work distribution function is of the form:

$$P(W,t) \overset{t \to \infty}{\sim} [4\pi \alpha \beta \upsilon^2 (t - t_0)]^{-\frac{1}{2}} \exp\left\{-\frac{[W - \alpha \beta \upsilon^2 (t - t_0)]^2}{4 \alpha \beta \upsilon^2 (t - t_0)}\right\}, \tag{23}$$

for *any* initial distribution $f(x_i, \dot{x}_i, t_o)$. We remark that $P(W,t)$ in the limit $t \to \infty$ depends only on $\alpha$ and $\beta$, but not on $m$, i.e. it holds both for the Langevin equation (1) with inertia $(m \neq 0)$ and without inertia $(m = 0$: the overdamped Langevin

---

[4] We note that both OM [4,5] and Bertini et al [15] only consider the optimal path of the Lagrangian $L$ i.e. $L_w$ with $\lambda = 0$.

equation).

From (23) one derives, again for any initial distribution, the asymptotic work fluctuation theorem:

$$\lim_{t \to \infty}\left[\frac{P(W,t)}{P(-W,t)}\right] = \exp W. \tag{24}$$

We note that (24) is independent of the frame, i.e. valid for $W(\{x_s\})$ of equation (15) for both $\vartheta = 0$ and $\vartheta = 1$.

Later we will generalize this sketch of the generalization of the OM theory from ES to NESS for this simple model to more general systems, and then determine in a more general way the work and heat needed to maintain a NESS.

### 6. Inertial effects for finite times – the critical mass

Although the asymptotic fluctuation relation for $P(W,t)$ is independent of inertial effects, they dramatically appear for finite times. It is convenient to introduce in this context a fluctuation function $G(W,t)$ defined by[5]:

$$G(W,t) = \frac{\partial}{\partial W}\left[\ln\frac{P(W,t)}{P(-W,t)}\right], \tag{25}$$

with, for both frames, $\vartheta = 0$ and $\vartheta = 1$, a simple asymptotic behavior: $\lim_{t \to \infty} G(t) = 1$.

Since the behavior of $G(W,t)$ for finite $t$ depends on the initial condition we choose in order to obtain concrete results, a NESS initial condition:

$$f(x_0, \dot{x}_0, t) = \left(\frac{\beta}{2\pi}\right)(m\kappa)^{1/2} \exp\left\{-\beta\left[\frac{1}{2}m(\dot{x}_0 - \upsilon)^2 + \frac{1}{2}\kappa[x_0 - \upsilon(t_0 - \tau_r)]^2\right]\right\}. \tag{26}$$

Equation (26) is a Gaussian distribution for the initial $x_0$ and $\dot{x}_0$ for the particle, around their average values of $\upsilon(t_0 - \tau_r)$ and $\upsilon$, respectively, where the stationary position $\upsilon(t_0 - \tau_r)$ of the particle in the laboratory frame is used.

---

[5] This fluctuation function differs from those used before, where a different scaling was used (cf. [16]). The reason is that (25) is adapted to the Gaussian form of the distribution function $G(W,t)$ [2] and provides more informative expressions here.

Evaluating $P(W,t)$ for the initial condition (26), a simple explicit form for $G(t)$ can be obtained [2], which is independent of $\upsilon$ and $\beta$, but does depend on $\vartheta$, i.e., on the frame. In addition, $G(t) > 0$ for $t > 0$ (cf figs 2 and 3).

A short discussion of the relaxation of $G(t)$ in the laboratory frame and the comoving frame to its asymptotic value 1 follows, which is illustrated in figs 2 and 3.

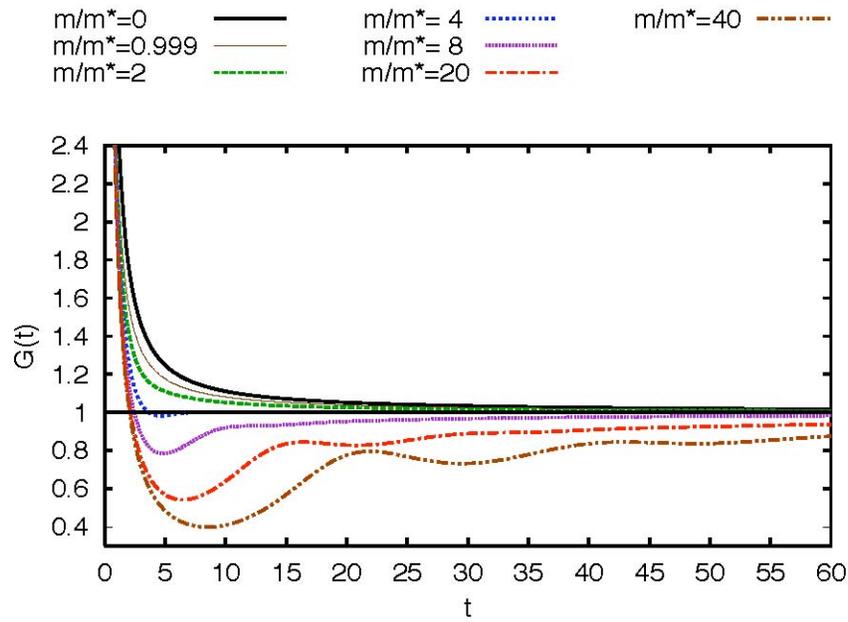

**Figure 2** The fluctuation function $G(t)$ as a function of time $t$ for the work done on the dragged Brownian particle in the laboratory frame $(\vartheta = 1)$ for a NESS initial condition, for $t \in [0,66]$. The colored lines in the figure correspond to parameter values of the scaled mass $m/m^*$ of the Brownian particle for $0 \le m/m^* \le 40$. In this figure we used $\alpha = \kappa = 1$ (so that $\tau_r = 1$ is a unit of time and $m^* = \frac{1}{4}$). The onset of propagation for $m > m^*$ is clearly visible.

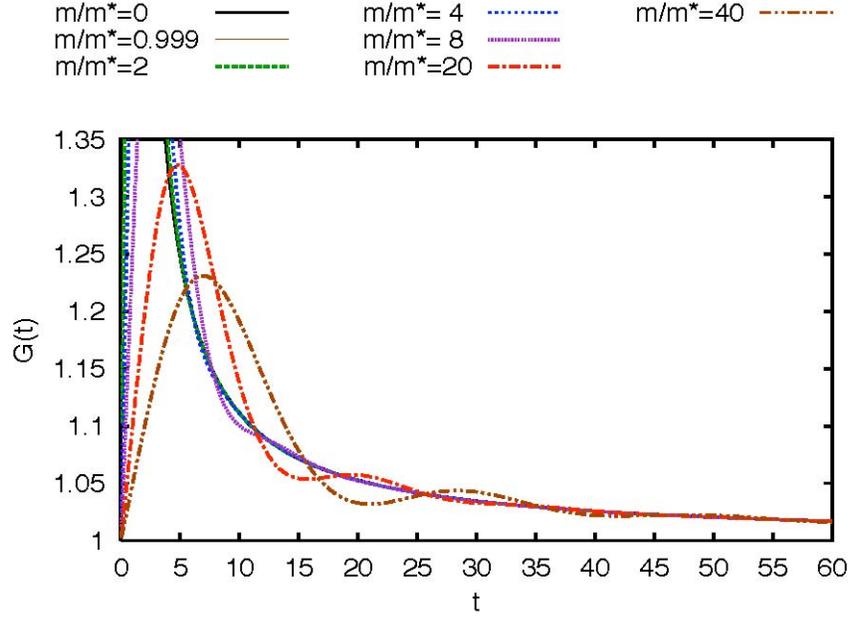

**Figure 3** The fluctuation function $G(t)$ as a function of time for the work done in the comoving frame $(\vartheta = 0)$. The depiction of this figure is the same as of figure 2.

A first approximation to $G(t)$ for large $t$ is [2]:

$$G(t) \overset{t \to \infty}{\sim} 1 + \frac{\tau_r - \tau_m \vartheta^2}{t - t_0 - \tau_r + \tau_m \vartheta^2}, \qquad (27)$$

i.e. for $\theta = 1$ (laboratory frame):

$$G(t) \overset{t \to \infty}{\sim} 1 + \frac{\tau_r - \tau_m}{t - t_0 - (\tau_r - \tau_m)} = 1 - \frac{1}{1 - \frac{t - t_0}{\tau_r - \tau_m}}. \qquad (27a)$$

while for $\vartheta = 1$ (comoving frame):

$$G(t) \overset{t \to \infty}{\sim} 1 + \frac{\tau_r}{t - t_0 - \tau_r}, \qquad (27b)$$

Although (27) is only a first approximation to the asymptotic form of $G(t)$, it includes an important *inertial* contribution to $G(t)$ via $\tau_m$ in the laboratory frame, which is absent in the comoving frame due to the d'Alembert-like force. Furthermore, in the laboratory frame, $G(t) \overset{t \to \infty}{\sim} 1 \pm O\left(\frac{\tau}{t}\right)$, depending on whether $\tau = \tau_r \gtreqless \tau_m$, respectively.

When $\tau_r = \tau_m$ or $m = \dfrac{\alpha^2}{\kappa}$, a higher approximation to $G(t)$ is needed to obtain the approach to its asymptotic value 1.

More importantly, there is a critical mass $m^*$, such that for $m > m^*$, $G(t)$ exhibits an oscillating behavior in time when approaching its asymptotic value 1, in the laboratory as well as in the comoving frame. This oscillating behavior is a direct manifestation of the general form solution of (20) for $\tilde{x}_s^*$ as a superposition of four terms of the form $\exp(v_j s)$, with $v_j$ given by equation (21), since all the $v_j$ become complex when $4\tau_m > \tau_r$ or for $m > m^* = \dfrac{\alpha^2}{4\kappa}$. It appears indeed numerically that at $m = m^*$ a "dynamical phase transition" takes place (cf figs 2 and 3).

For $m > m^*$ the position $\tilde{x}_s^* = x_s^* + (1-2\lambda)\upsilon\tau_r$, i.e. the position of the Brownian particle, oscillates with a period:

$$T_m = 2\pi\sqrt{\dfrac{m}{\kappa}}\ (1 - \dfrac{m^*}{m})^{-\tfrac{1}{2}} \qquad (28)$$

corresponding to a frequency $\omega = \sqrt{(4\tau_m \tau_r^{-1} - 1)}/2\tau_m = |\mathrm{Im}\, v_\pm|$ with (21). Clearly the existence of two time scales, one of which is $\tau_m$ and $\tau_r$ are necessary for the occurrence of time oscillatory behavior in both frames. We note that in the overdamped case, where $\tau_m = 0$, no oscillations occur.

We emphasize that the finite time behavior of the work fluctuation function $G(t)$ discussed here for the NESS system is completely *absent* in the ES. In fact, in the NESS the approach of $G(t)$ to its asymptotic behavior depends dramatically on the frame considered. This is another manifestation of the lack of Galilei invariance of the Langevin equations (3) and (5).

There are many other interesting features in figures 2 and 3, illustrating the different behavior of $G(t)$ in the laboratory and the comoving frames, which could all be observed, but for a detailed discussion of which we refer to the literature [2].

Finally, in view of the analogy of this model with an electric circuit [3], a critical value $L^*$ of the self-inductance $L$ could play the same role in the latter as the critical mass $m^*$ in the former. This would lead to a similar behavior of the work fluctuation function $G(t)$ as discussed in section 6.

### 7. Thermodynamics for a NESS

In this section we give a more general presentation of the generalization of the OM theory to systems in a NESS, than just for the dragged Brownian particle system.

Defining the heat $Q_{NESS}$ and the work $W_{NESS}$ in a NESS as those needed to maintain the NESS, a formal analogy to the first and second laws of thermodynamics in an ES can be formulated for the NESS. In the following all subscripts NESS will be dropped to simplify the notation.

The existence of NESS thermodynamic laws should not be surprising since they merely express for the NESS the conservation of energy (NESS first law) and the existence of a positive entropy production (NESS second law), as they do in an ES. No statement is made about the entropy of a NESS itself.

In order to discuss a formulation of the laws of thermodynamics in a NESS, we first need definitions of the NESS heat $Q$ and the NESS work $W$ for a system in such a state. Due to the very nature of the NESS the proper choice of $Q$ and $W$ cannot be made without specifying the coupling of the system to other systems, with which it interacts i.e its physical nonequilibrium parameters, which will be collectively indicated by μ.[6] This is an essential difference between NESS and ES thermodynamics, where in the latter this ambiguity does not occur.

The origin of the ambiguity is that heat and work in the generalized OM theory are defined via paths in a function space [3]. In fact, they are defined in terms of a forward and a corresponding backward path, where the latter is obtained by a time-reversal procedure. It is this procedure which is unique in an ES, but not a priori in a NESS. While in an ES, the *only* time reversal is, apart from $s \to -s,$ that of the "internal" particle velocity $\dot{x}_s$ into $-\dot{x}_s$, there is an additional option in a NESS to change the sign of the nonequilibrium parameter $\mu$. In the dragged Brownian particle system, for example, one can or cannot change the sign of the velocity $\upsilon$.

---

[6] A nonequilibrium parameter μ need not be a dynamical quantity, but can also be a temperature difference, as in two examples in [3]. We also note that for $\mu = 0$, the OM results for a system in an ES are regained. In general, like the coordinates $x_s$, the nonequilibrium parameter μ can be vectors, whose components indicate the number of degrees of freedom of the system of interest and of the number of nonequilibrium parameters, respectively.

To discuss this time reversal ambiguity arising in a NESS mathematically, it is convenient to introduce a time reversal operator $\hat{I}_\pm$ defined for any functional $\mathcal{F}(\{x_s\};\mu)$ by:

$$\hat{I}_\pm \mathcal{F}(\{x_s\};\mu) = \mathcal{F}(\{x_{t+t_0-s}\};\pm\mu) \text{ with } s \in [t_0,t], \tag{27}$$

where $\mu$ characterizes the nonequilibrium parameters, which specify the NESS system.

Thus under this time reversal operator the motion of the (Brownian) particle on a *forward* path $\{x_s\}_{s\in(t_0,t)}$ is transformed into a motion along a *backward* (time-reversed) path $\{x_{t+t_0-s}\}_{s\in[t_0,t]}$ with the same path geometry but initial and final positions given by $x_t$ at time $t_0$ and $x_{t_0}$ at time $t$, respectively, while the corresponding positions on the forward path were $x_{t_0}$ at time $t_0$ and $x_t$ at time $t$, respectively.

For technical reasons we replace from now on, without loss of generality, the origin of time at $\frac{(t_0+t)}{2}$ so that $t_0 \to -t$ and $t \to t$ (cf fig 4). Note that the initial and final positions on the backward path are identical to the final and initial positions, on the forward path, respectively. The time reversal for the "*internal*" motion of the particle of interest, $x_s$, is indicated by the hat (^) on the operator $\hat{I}_\pm$. On the other hand the time reversal procedure associated with the nonequilibrium parameters, indicated collectively by $\mu$, is indicated by the subscripts $\pm$, so that one has either $\hat{I}_-$ or $\hat{I}_+$. These operators do or do not change the *sign* of the components of the nonequilibrium parameter $\mu$ under a time reversal operation, i.e. in the dragged particle case from $\upsilon$ to $-\upsilon$ or from $\upsilon$ to $\upsilon$, respectively.

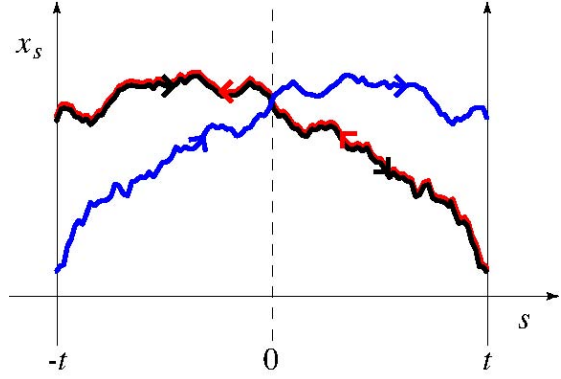

**Figure 4** A forward path from an initial state $x_{-t}$ at time $-t$ to a final state $x_t$ at time $t$ (black line and arrows) and its corresponding backward path from an initial state $x_t$ at $t$ to a path state $x_{-t}$ at $-t$, with opposite time direction and velocities (red line and arrows) (cf (28b)). Equivalent backward path obtained from the same forward path by visiting the states in the opposite order $x_t$ at $-t$ to $x_{-t}$ at $t$ (blue line and arrows) (cf (28a)).

$\hat{I}_\pm$ has two important properties:

1. $\hat{I}_\pm^2 = 1$

2. $\hat{I}_\pm \int_{-t}^{t} ds\, f(X_s; \mu) = \int_{-t}^{t} ds\, f(\ddot{x}_{-s}, \dot{x}_{-s}, x_{-s}, s; \pm\mu),$ (28a)

$$= \int_{-t}^{t} ds\, f(\ddot{x}_s, -\dot{x}_s, x_s, -s; \pm\mu)$$ (28b)

$$= \int_{-t}^{t} ds\, f(X_{-s}; \pm\mu)$$

for any function $f(X_s; \mu)$ with $X_s \equiv (\ddot{x}_s, \dot{x}_s, x_s, s)$ and $X_{-s} \equiv (\ddot{x}_s, -\dot{x}_s, x_s, -s)$.

We note that the possibility of a sign change of $\mu$ is directly related to the more complex structure of a NESS than an ES.

The central quantity in the generalized OM theory of a NESS is the heat, which is directly related to the stochastic Lagrangian on which the theory is based. This also allows a direct derivation of *the second law for NESS thermodynamics*. We proceed in three steps: derivation of (a) the entropy production rate, (b) the second law and (c) the heat associated with the NESS.

Ad (a). We separate the stochastic Lagrangian (cf (11)) into a time-reversal symmetric and a time-reversal antisymmetric part:

$$L(X_s;\mu) = -\frac{1}{2k_B}\left[\Phi_\pm(X_s;\mu) - \dot{S}_\pm(X_s;\mu)\right], \quad (29)$$

with

$$\Phi_\pm(X_s;\mu) = -k_B\left[L(X_s;\mu) + L(X_{-s};\pm\mu)\right], \quad (30)$$

and

$$\dot{S}_\pm(X_s;\mu) = k_B\left[L(X_s;\mu) - L(X_{-s};\pm\mu)\right], \quad (31)$$

which correspond to the symmetric and antisymmetric parts with respect to time reversal of the Lagrangian $L$. Here, as everywhere in this paper, the + and – signs on both sides of an equation correspond to each other.

A major step in the (generalized) OM theory is now to identify the physical entropy production rate of the system of interest with the *antisymmetric* part of the Lagrangian $L$ with respect to time reversal i.e. with $\dot{S}_\pm(X_s;\mu)$.

Ad (b). The second law of thermodynamics follows now in two steps.

1. This law is expressed in terms of the most probable or average positions $\langle x_s \rangle$, given by the solutions of the *average* Langevin equation, using $\langle \zeta_s \rangle = 0$.[7]

2. One can then prove the positivity of the entropy production rate $\dot{S}_\pm(\langle X_s \rangle;\mu)$ by:

$$\dot{S}_\pm(\langle X_s \rangle;\mu) = \Phi_\pm(\langle X_s \rangle;\mu) > 0, \quad (32)$$

using that the dissipation function $\Phi_\pm > 0$ for all paths $X_s$ in a NESS and (8). For further details we refer to [3][8,9]. The formulation of the second law here in terms of $\langle X_s \rangle$ instead of $X_s$ is a generalization of OM's procedure for the ES, where $\mu = 0$. The

---

[7] Alternatively, by the condition $L(X_s;\mu)$ = maximum.

[8] For the dragged Brownian particle model, $\mu = \upsilon$ and

$$\Phi_-(\langle X_s \rangle;\upsilon) = \frac{\alpha}{2T}\langle \dot{x}_s \rangle^2 + \frac{1}{2\alpha T}\left[\kappa\left(\langle x_s \rangle - \upsilon s\right) - m\langle \ddot{x}_s \rangle\right]^2,$$ since in this model $\hat{I}_-\left(\upsilon s\right) = \upsilon s$.

[9] There is, in general, an additional contribution to $\Phi_\pm$ or $\dot{S}_\pm$, which can be ignored for all eight models considered in [3].

appearance of $\langle X_s \rangle$ can perhaps be understood on the basis that the second law is a macroscopic law, formulated in terms of macroscopic, i.e. average, quantities $\langle X_s \rangle$ rather than microscopic quantities $X_s$.

Ad (c). Finally the heat for the NESS can be defined in terms of the entropy production rate by:

$$Q_\pm(\{x_s\};\mu) = T \int_{-t}^{t} ds \; \dot{S}_\pm(X_s;\mu). \tag{33}$$

This shows that the heat is directly related to the time-antisymmetric part of the Lagrangian (33) (cf (35)), or:

$$\hat{I}_\pm Q_\pm(\{x_s\};\mu) = -Q_\pm(\{x_s\};\mu). \tag{34}$$

A direct connection of $Q_\pm(\{x_s\}, m)$ with the time-irreversible part of the basic Lagrangian $L(\{x_s\}, \mu)$ of the generalized OM theory is with (34):

$$Q_\pm(\{x_s\};\mu) = \beta^{-1} \int_{-t}^{t} ds [L(X_s;\mu) - L(X_{-s}; \pm\mu)]$$

$$= \beta^{-1} \ln \frac{P(\{x_s\};\mu)}{\hat{I}_\pm P(\{x_s\};\mu)}. \tag{35}$$

Here $P_\pm(\{x_s\};\mu)$ is proportional to the probability for a path $\{x_s\}$ for the nonequilibrium parameters $\mu$ (cf (13):

$$P(\{x_s\};\mu) \sim \exp \int_{-t}^{t} ds \; L(X_s;\mu). \tag{36}$$

This shows that it is the different behavior of $P(\{x_s\};\mu)$ on the forward path and the backward path which is responsible for the heat to maintain the NESS [4].

In order to obtain the *first law of NESS thermodynamics* in a concrete mathematical form, one has to introduce the internal energy and the energy conservation law. The internal energy $E_\pm(x_s, \dot{x}_s; \mu)$ is given as a sum of a kinetic and a potential energy contribution and is *assumed* to be time reversal invariant, leading to an internal energy difference between the final and the initial time:

$$\Delta E_\pm = E_\pm(x_t, \dot{x}_t; \mu) - E_\pm(x_{-t}, \dot{x}_{-t}; \mu), \tag{37a}$$

and

$$\hat{I}_\pm \Delta E_\pm = -\Delta E_\pm. \tag{37b}$$

We define the work $W_\pm$ done to maintain the NESS as the sum of the heat produced and subsequently removed from the system and the change of the internal energy of the system i.e. (cf [3]):

$$W_\pm(\{x_s\}; \mu) = Q_\pm(\{x_s\}; \mu) + \Delta E_\pm. \tag{38}$$

We note that the work $W_\pm(\{x_s\}; \mu)$ is, like the heat and the internal energy, antisymmetric under time reversal of the external nonequilibrium parameters $\mu$:

$$\hat{I}_\pm W_\pm(\{x_s\}; \mu) = -W_\pm(\{x_s\}; \mu). \tag{39}$$

This is a "local" formulation of the first law of NESS thermodynamics for each path $\{x_s\}$[10] formally identical to the first law of equilibrium the thermodynamics. These "local" formulations of the first and second law for a NESS can be generalized to a "global" formulation of the first law of NESS by carrying out a functional average $\langle\langle \cdots \rangle\rangle_t$ (cf (14)) over a region in function space between any initial and final state at $-t$ and $t$, respectively. This functional average appears as the analogue for systems with a stochastic dynamics of a phase space average for systems with a deterministic dynamics.

Because of the time reversal ambiguity, there will be for any number of nonequilibrium parameters, in principle, as many formulations of the first two laws of NESS thermodynamics, as there are external nonequilibrium parameters, indicated by the number of components of the vector $\mu$.

This ambiguity therefore leads to the fundamental question, which of the external nonequilibrium parameters $\mu$ or $-\mu$ has to be chosen to obtain the correct physical work and heat to maintain the NESS. The two laws of NESS thermodynamics alone do not give a resolution of this ambiguity.

---

[10] We emphasize that the NESS thermodynamics formulated in (32) and (38) should be distinguished from the well-established field of Irreversible Thermodynamics (or Thermodynamics of Irreversible Processes), as discussed e.g. in [17] and which is based on two earlier papers by Onsager [18] than [4,5].

It appears that one can only make a unique physically acceptable choice *a posteriori* on physical grounds *after* having specified *a priori* the NESS system concretely i.e. its nonequilibrium parameters. One can expect, as is borne out by the examples we have studied, that different models will require different choices of $\hat{I}_+$ or $\hat{I}_-$, to obtain a physically acceptable theory.

It appears that a resolution[11] of this ambiguity can be obtained by taking into account the parity with respect to time reversal of the nonequilibrium parameters, which characterize the system in the NESS. Then it should follow that the (proper) physical work, heat and internal energy obtained satisfy the three conditions of time reversibility of the internal energy, positivity of the average work done on and the average heat removed from the system in a NESS.

This has been illustrated in [3] for eight different models, with a variety of nonequilibrium parameters. One of these models is the dragged particle model used as an illustration in this paper. Additional models include electrical circuits, a driven torsion pendulum, and an energy current generated by a temperature difference. These models are devised to illustrate different resolutions of the ambiguities for the choice of the proper physical work, heat, and internal energy associated with them. In fact, four require $\hat{I}_-$ and four $\hat{I}_+$, respectively. We note that results for the models considered here have been or can be verified experimentally.

## 8. Transient Work Fluctuation Relations and Nonequilibrium Detailed Balance Relations

Transient fluctuation relations [19] can be derived from nonequilibrium detailed balance relations. We will illustrate this on the dragged Brownian particle model and refer for a more general formulation and connections with the literature to [1-3, 19, 20,].

We first describe two nonequilibrium detailed balance relations, one for the mechanical (NESS thermodynamic) work *W* done on the system to maintain it in a NESS and another for the friction work *R* done by the Brownian particle to overcome the friction in the fluid.

---

[11] This resolution is not incorporated in the papers [1-3].

For the mechanical work $W$ in the laboratory frame, defined by (15a), the following nonequilibrium detailed balance relation holds (cf [1]):

$$\exp(-\beta W(\{x_s\};\upsilon)) \, \exp\int_{t_0}^{t} ds \, L(X_s;\upsilon) \, f_{eq}(\dot{x}_0, x_0)$$

$$= \exp\int_{t_0}^{t} L(X_{-s};-\upsilon) \, f_{eq}(\dot{x}_t, x_t), \tag{40a}$$

while for the friction work $R$, in the commoving frame, defined below by (41), one has:

$$\exp\left(-\beta R(y_t, y_0;\upsilon)\right) \exp\int_{t_0}^{t} ds \, L(Y_s;\upsilon) \, f_{eq}(\dot{y}_0, y_0)$$

$$= \exp\int_{t_0}^{t} ds \, L(Y_{-s};\upsilon) \, f_{eq}(\dot{y}_t, y_t). \tag{40b}$$

On the left hand side (l.h.s.) of both equations is a Boltzmann factor, which incorporates the mechanical work $W$ or the frictional work $R$, respectively. Since both vanish in equilibrium, when the nonequilibrium parameter $\upsilon = 0$, these relations reduce then to the usual equilibrium detailed balance relations. While in (40a) the mechanical work is given by an integral over the full path $\{x_s\}_{s\in[t_0,t]}$ (cf (15a)), in (40b) the friction work is given by a boundary term:

$$R(y_t, y_0;\upsilon) = \int_{t_0}^{t} ds \, (-\alpha \dot{y}_s) \, \upsilon = -\alpha\upsilon \, (y_t - y_{t_0}). \tag{41}$$

The other two factors on the l.h.s. and r.h.s. of (40a,b) give the probabilities associated with the appropriate forward and backward paths in function space and the corresponding initial states for these two paths, respectively.

The transient fluctuation relations for $W$ and $R$, which can be derived from (40a) and (40b), respectively for an initial equilibrium distribution, using (11) and (15a) and read [1]:

$$\frac{P(W,t)}{P(-W,t)} = \exp W, \tag{42a}$$

for $W$ and similarly for $R$:

$$\frac{P(R,t)}{P(-R,t)} = \exp R. \tag{42b}$$

Here $W$ and $R$ are both dimensionless work (cf (17)).

Comparing the asymptotic fluctuation theorem (24) for $W$ and the transient fluctuation theorems for $W$ and $R$, (42), we note that although in both cases work is involved, for the mechanical work $W$ the appropriate operator is $\hat{I}_-$ (i.e. $\dot{x}_s \to -\dot{x}_s$ and $\upsilon \to -\upsilon$), while for the friction work $R$, on the other hand, the operator is $\hat{I}_+$ ((i.e. $\dot{y}_s \to -\dot{y}_s$ and $\upsilon \to \upsilon$) has to be used in the Lagrangians (40a) and (40b), respectively, to obtain (41a) and (41b), respectively.

We illustrate this further for $R$ for the overdamped case in the comoving frame, i.e., (6) with $m = 0$, for the dragged Brownian particle model by choosing an initial condition of the form:

$$f(y_0, t_0) = f_{eq}(y_0 + \upsilon \tau_r \phi), \tag{43a}$$

with a constant parameter $0 \leq \phi \leq 1$ and

$$f_{eq}(y) = \left(\frac{\kappa\beta}{2\pi}\right)^{\frac{1}{2}} \exp\left[-\frac{\beta}{2}\kappa y^2\right]. \tag{43b}$$

Using the probability distribution function $P(R,t)$ for this initial condition [1] gives the fluctuation relation:

$$\frac{P(R,t)}{P(-R,t)} = \exp[(1-\phi)R] \tag{44}$$

We note that (44) only has the form of a transient fluctuation relation for $\phi = 0$ (cf (42b)), i.e., for an initial equilibrium distribution function, while for $\phi = 1$, i.e. for an initial nonequilibrium steady state distribution, $P(R,t)$ is Gaussian, with a peak at $R=0$ and $P(R,t) = P(-R,t)$ so that their ratio is 1.

We emphasize that the above mentioned difference in the definitions of $W$ and $R$ - an integral as opposed to a boundary term – is ultimately responsible for the fact that, while there are both a transient (42a) and an asymptotic fluctuation relation (24) for $W$, there is *only* a transient fluctuation for $R$ (42b) and no asymptotic fluctuation relation, since $R$ never looses its memory of the initial state.

One can summarize the difference between the transient and the asymptotic fluctuation relations in that the former holds for *all* times $t$, but *only* for an equilibrium initial condition, while the latter holds for *all* initial conditions, but *only* for $t \to \infty$ (see also [21]).

9. Comments and Open Questions

1. The basic assumption of this paper, as to the proper definition of heat and work for a system in a NESS, bears some semblance with a case considered by Landau and Lifshitz [22]. They considered a complex thermally isolated system, not in thermal equilibrium, consisting of a number of (sub) systems which interact with each other and can do work on external objects. The approach of such a system to equilibrium, as well as the final equilibrium state itself, are then non-unique and can only be determined if one specifies the complex system concretely, i.e. identifies its nonequilibrium parameters, characterizing the couplings between all the subsystems and to the external objects. They say: "Here one is only interested in the work produced due to the fact that the system is not in equilibrium. This means that we must ignore the work done [e.g.] by a general expansion of the system, work which could also have been done by the system in an equilibrium state." This last condition also pertains to this paper when defining the heat and the work in a NESS.

2. In addition to the ambiguity related to the proper choice of the sign of the nonequilibrium parameters, there is a second ambiguity, similar, but different from, that which exists in the thermodynamics of a system in an ES. There, as far as the first law of thermodynamics is concerned, one can add to the heat and the work the same constant without violating the (energy conservation) law. Here, for a system in a NESS, the same ambiguity exists with respect to the separation of the (NESS) heat into a sum of (NESS) work and (NESS) internal energy difference, which can also be done only up to a common constant, respectively. This can e.g. be illustrated in the dragged Brownian particle model, where the difference in the work in the laboratory frame and the comoving frame equals the difference in the internal energy (cf 15c,d), while the heat is the same in the two frames.

3. The derivation of the Second Law of NESS Thermodynamics presented here is incomplete for two reasons. First, because the forces acting on the system have to be linear. This makes a generalization of the present theory to the non-linear case particularly important. In the dragged Brownian particle model this means a non-linear force in the particle position (cf (8)) and/or e.g. a non-linear dependence of the friction of the particle on the dragging velocity.

  4. We have restricted ourselves in this paper to a discussion of properties of the work in a system in a NESS. For a discussion of the heat in such a system, we refer to the original papers [1,3].

  5. We note that the resolution of the ambiguity of the choice of the proper backward path for a given forward path mentioned in Section 7 yields the heat, work and energy needed to maintain the system in a NESS. These same quantities also satisfy the laws of NESS thermodynamics. However, as the dragged Brownian particle model shows, one can obtain in addition to the mechanical work, which has all the above properties, an other physical quantity in this model: the friction work, which is also a measureable physical quantity but has different properties (e.g. different fluctuation relations) than the mechanical work. It is unclear, at present, how general this consequence of the sign ambiguity of the nonequilibrium parameters found in the dragged particle model is.

  6. From a general point of view, one can say that the behavior of the systems in a NESS considered here, is determined by that of its Langragian $L(X_s;\mu)$ as a function of the paths $\{x_s\}$ and the nonequilibrium parameter(s) $\mu$, under the action of the time reversal operator $\hat{I}_\pm$. When the probability of a forward path differs from that of the corresponding backward path in function space, dissipation occurs, due to the entropy production in the NESS (cf (33) and (35))[12].

On the other hand, the behavior of a system in an ES, where there are no nonequilibrium parameters $\mu$ and only the velocity $\dot{x}_s$ (and the time $s$) will change sign upon time reversal. Although there is then still no equal probability for a forward and a backward path (e.g., cf (11) and (13)) no entropy production will occur on average.

  7. In this paper we confined ourselves to "systems of interest," whose NESS properties are due to them being part of a larger system, which is characterized by

---

[12] The same is true for a deterministic system in phase space, where the phase space contraction represents the dissipation or entropy production [8, 21].

*physical* nonequilibrium parameters $\mu$. In fact, for the class systems of interest considered in [3], these parameters had a distinct parity with respect to time reversal. For the OM path integral approach this is relevant for the unique determination of the proper backward path corresponding to a given forward path of the system of interest in a NESS. However, there are other nonequilibrium parameters as, e.g., the friction coefficient $\alpha$ in the dragged Brownian particle model, or the resistance in an electric circuit, which represent the coupling to a heat bath surrounding the system of interest. They were not included in the nonequilibrium parameters $\mu$. This is because there is here no time reversal ambiguity, since the second law requires that both are >0. It is not clear at present how, in general, the choice of unique corresponding forward and backward paths for any system in a NESS should be made, without a more detailed classification of their possible compositions.

8. In their 1953 paper [4], Onsager and Machlup mention possible generalizations of their approach to other systems than those in an ES. They say: "The extension to [other systems] ..... usually involves simply a change of language ..... But again, the extension to open systems (and steady states) is purely formal". In our generalization of the OM approach to systems in a NESS this appears not to be so.

Acknowledgements

The author is indebted to Dr. T. Taniguchi and Dr. H. Touchette for many helpful discussions and Mr. J Alonzo for assistance in the preparation of this manuscript. He also gratefully acknowledges financial support of the Mathematical Physics program of the National Science Foundation under grant PHY-50315.